Thierry DELORT                                            10th November 2011
9 rue MALTE BRUN
75020
PARIS
France
Email :tdelort@yahoo.fr

Title: Interpretation of the supraluminous velocity of neutrinos by a Theory of Ether.version 2

Résumé:

We exposed between the years 2000 to 2011 a very general theory of Ether, giving an interpretation of all the main experiments connected to Special and General Relativity and Cosmology. This theory of Ether was most of time even not considered, because it was contrary to Special Relativity. But a French team of physicists (led by Dario Autiero) recently realized an experiment, whose the result was contrary to Special Relativity, because it implied that a particle could go faster that the light. We could suppose that this experiment was not good and due to an experimental error , but another solution is that Special Relativity is not good. And if it is the case, the modern Theory of Ether becomes very interesting, being the only one complete alternative theory to Special Relativity. In this article, we are going to give the interpretation of the experiment of Dario Autiero by the modern Theory of Ether.

1.INTRODUCTION

The team of physicists led by Dario Autiero just realized an experiment showing that some neutrinos can go faster that the light. This experiment contradicts Special Relativity. Some physicists pretend that the result of this experiment is not correct, being due to experimental error, and keep admitting Special Relativity. But it is also very likely that this experiment be correct (it has been realized very carefully), and that Special Relativity is itself wrong. In this case, the only alternative to Special Relativity is the modern Theory of Ether. Because not only this theory permits to interpret all main experiments connected to Special and General Relativity as well as Cosmology, but also because it permits to give a very simple and amazing explanation to the experimental result of the team of Darion Autiero. This modern Theory of Ether has been published in 7 articles written in English [1][2][3][4][5][7][8] ,all those articles being modified and improved in a French version published in the book [6]. The part of this Theory of Ether corresponding to Special Relativity is mainly exposed in the article [1], and in the improved version of this article [6].
In this article, we just expose the interpretation by the Theory of Ether of the experiment of D.Autiero, and we remind explicitly all the elements of the Theory of Ether that we will use in order to interpret this experiment.

So in this article we will use only 3 elements of the modern Theory of Ether corresponding to Special Relativity:

The first element is the existence of an absolute Referential, called "Ether".

The 2nd element is that the transformations between the Ether (absolute Referential R) and a Galilean Referential R′ are, with evident and conventional notations and hypothesis,(The Galilean Referential being driven with a velocity v relative to the Ether, the axis OX and OøXø coinciding and OøYø and OøZø being respectively parallel to OY and OZ):

$$X' = \frac{X - vT}{\sqrt{1 - v^2/c^2}}$$
Yø=Y
Zø=Z
$$T' = T\sqrt{1 - v^2/c^2} \quad (1)$$

Times and lengths in Rø are measured by standard clocks and rules, at rest relative to Rø
We also remarked that if we only delay the clocks of Rø of $vXø/c^2$, (keeping the same axis and the same spatial measurement), then we obtain a Lorentz Referential Rö, the transformations between R and Rö being:

$$X'' = X' = \frac{X - vT}{\sqrt{1 - v^2/c^2}}$$
Yö=Yø=Y
Zö=Zø=Z

$$T'' = T' - vX'/c^2 = \frac{T - vX/c^2}{\sqrt{1 - v^2/c^2}} \quad (2)$$

Considering the 2 previous transformations, we have 2 fundamental remarks A and B:
A. An event (X,Y,Z,cT) has the same spatial coordinates in Rø and Rö. A fixed point of Rø is also a fixed point of Rö.
B. In a fixed point of Rø (or of Rö), the time interval between 2 events is the same measured in Rø as the one measured in Rö: Tø= Tö because Xø=0.

(Using also those transformations we obtain in the article [1][6] that trajectories of particles can be calculated in Rø or in Rö, and also that usual experiments in electromagnetism cannot reveal the velocity of the earth relative to the Ether)

The 3$^{rd}$ element of the Theory of Ether that we will use is that in this theory, there is not a limit to the theoretical velocity of a particle relative to the Ether (except if it owns a positive mass) or to a transmission of information.
Moreover, we obtain in Cosmology that the velocity v of our Galilean Referential relative to the Ether, verifies v/c<<1 (In particular according to the Theory of Ether, v is our own velocity relative to the Referential in which fossil radiation is isotropic. So v is of the order of $10^{-3}c$).

2. THEORY

We are going to justify in this section why all happens (in the experiment of D.Autiero) as if neutrinos were driven in our Referential with a velocity c(1+ ) (with <<1), and the photons driven with a velocity c.

The 2 simplest possible theoretical explanations are the following ones:

1st possible theoretical explanation:

Because there is no limit to the theoretical velocity of the neutrinos relative to the ether, we can assume that they are driven with a constant velocity $c(1+\varepsilon)$ with $\varepsilon \ll 1$. $\varepsilon$ should depend on the energy (absolute, meaning relative to the Ether) and on the nature of the neutrino.

We consider the Galilean Referential Rø in which we live and its associate Lorentz Referential Rö that we defined in 1.INTRODUCTION.

We obtain the transformations of velocities between Rö and the Ether, with evident notations:

$$V_X" = \frac{V_X - v}{1 - vV_X/c^2}$$

$$V_Y" = \frac{V_Y}{1 - vV_X/c^2}$$

$$V_Z" = \frac{V_Z}{1 - vV_X/c^2} \qquad (3)$$

So let us suppose that the absolute velocity (meaning velocity relative to the Ether) of a neutrino is $v_n = (1+\varepsilon)(c_X, c_Y, c_Z)$. So $(c_X, c_Y, c_Z)$ are the coordinates of a virtual photon, with $c_X^2 + c_Y^2 + c_Z^2 = c^2$.

Then using the previous transformations (3) and the fact that $v \ll c$ and $\varepsilon \ll 1$, we obtain that the velocity of the neutrino in Rö is $v_n = (1+\varepsilon)(c_X ö, c_Y ö, c_Z ö)$, with an error of the order of $v\varepsilon/c$, $(c_X ö, c_Y ö, c_Z ö)$ being the coordinates of the virtual photon in Rö. Let us show more clearly how we obtain this result:

According to the equation (3):

$$c_X" = \frac{c_X - v}{1 - vc_X/c^2}$$

$$c_Y" = \frac{c_Y}{1 - vc_X/c^2}$$

$$c_Z" = \frac{c_Z}{1 - vc_X/c^2} \qquad (4)$$

Also according to the equation (3) and the fact that in the Ether $v_n = (1+\varepsilon)(c_X, c_Y, c_Z)$, $(v_{nX}ö, v_{nY}ö, v_{nZ}ö)$ being the velocity of the neutrino in Rö:

$$v_{nX}" = \frac{(1+\varepsilon)c_X - v}{1 - (1+\varepsilon)vc_X/c^2}$$

$$v_{nY}" = \frac{(1+\varepsilon)c_Y}{1 - (1+\varepsilon)vc_X/c^2}$$

$$v_{nZ}" = \frac{(1+\varepsilon)c_Z}{1 - (1+\varepsilon)vc_X/c^2} \qquad (5)$$

Using the equations (5) and (4), we easily obtain that in $R\emptyset$, the velocity of a neutrino is $(1+ )(c_X ö, c_Y ö, c_Z ö)$ with an approximation of the order $v/c$.

This means that the velocity of the neutrino in $Rö$ is $v_n = c(1+ )$, with an error of the order $v/c$. (Because $c_X^2 ö + c_Y^2 ö + c_Z^2 ö = c^2$, $Rö$ being a Lorentz Referential).
For instance in the particular case, $c_X = c_Z = 0$, we obtain exactly with obvious notations $v_{nY} ö = V_{nY} = c(1+ )$.

According to the Remark B of the introduction, the departure of a neutrino and of a photon are simultaneous in both $R\emptyset$ and $Rö$, and also the delay of the photon relative to the neutrino can be calculated in $Rö$ as in $R\emptyset$

So we justified theoretically completely the experiment of the team of D.Auterio, meaning that in order to obtain the delay of photons relative to neutrinos, we just can consider that all happens as if photons were driven with a velocity c, and neutrinos were driven at a velocity $(1+ )c$ in our Referential.

2$^{nd}$ possible theoretical explanation:

So according to the first theoretical explanation, we only need to obtain the velocity of a neutrino in the absolute frame (Ether). It seems reasonable that it depend on energy. The simplest and more attractive possibility would be that we have a velocity c*, close to c but slightly superior to c such that we have:

$$E = \frac{mc*^2}{\sqrt{1 - V^2/c*^2}} \qquad (6)$$

The previous equation would be valid for neutral elementary particles (fermions) such as neutrinos. For all the other charged elementary particles, meaning in particular all quarks and also electrons and muons, the usual expression of the energy would be valid. Consequently the neutral particles constituted of quarks would also obey to the classical expression of energy.
We then obtain very simply the expression of the impulsion for neutrinos:

$$\mathbf{P} = \frac{m\mathbf{V}}{\sqrt{1 - V^2/c*^2}} \qquad (7)$$

So we keep the fundamental equation: $E^2 - P^2 c*^2 = mc*^2$.
According to Opera experiment, $(c*-c)/c$ is of the order of $10^{-4}$.
The previous theory is not possible in Relativity in which a particle cannot have a velocity superior to c.
The equation (6) justifies that the velocity increase with the energy, in agreement with Opera experiment and the observation of neutrinos emitted by a supernova. The fact that it be not valid for charged particles explains why we do not find supraluminous velocity for electrons and muons, despite that they be coupled with neutrinos.

3.DISCUSSION

So the modern Theory of Ether appears to be one among the very rare theories permitting to interpret the whole of experiments connected to Special Relativity as well as the experiment of D.Autiero. It is based on the fundamental fact that inside the Theory of Ether, there is not any limit velocity to a transmission of information.

We could wonder why electrons, that are leptons coupled to neutrinos, cannot reach supraluminous velocities. But the explanation is that the equation (6) is valid only for neutral elementary particles which is not the case for electrons.

According to Cohen-Glashow effect, a particle driven with a velocity superior to the velocity of the light radiates some energy. But this effect is predicted only if we assume the Relativity Theory, which is not the case in the Theory of Ether.

To end we could wonder why in the case of neutrinos emitted by a supernova, we observed a velocity of neutrinos much smaller than the velocity of neutrinos in the experiment of D.Autiero. ((v/c-1) is of the order $10^{-8}$ in a case and $10^{-5}$ in the $2^{nd}$ case). The explanation is that they do not have the same absolute energy. Comparing their energy could be a test to verify the $2^{nd}$ possible theoretical explanation.

4.CONCLUSION

So we showed that the experiment of Dario Autiero was not necessarily due to an experimental error, but that it could really contradict Special Relativity and prove its invalidity . Then it appears to be a very great element confirming the validity of the modern Theory of Ether. The main point is that according to the experiment of D.Autiero, it is possible for a particle to go faster than the light, which contradicts Special Relativity and is in agreement with the modern Theory of Ether.  We remind that up to date, Theory of Ether and Special Relativity have the same prediction for all classical laboratory experiments, except for those linked to Quantum Physics (some of them contradicting also Special Relativity (Quantum intrication..)), consequently the experiment of D.Autiero appears to be the main laboratory experiment contradicting Special Relativity.

References:
1.T.Delort, Theory of Ether, Physics Essays 13,4 (2000)
2.T.Delort, Application of the Theory of Ether, Physics Essays 17,3 (2004)
3.T.Delort, Complements of the Theory of Ether, Physics Essays 18,2 (2005)
4.T.Delort,Theory of Ether with Gravitation, Physics Essays 18,1 (2005)
5.T.Delort, Follow up on the Theory of Ether, Physics Essays 20,3 (2007)
6.T.Delort, Théories dør 2ième edition, Edition Books on Demand, 2011 (website www.theoriesdor.com,with reactualized versions of the articles 1. to 5).
7.T.Delort, Elements of Cosmology in the Theory of Ether,PIRT 2004
8.T.Delort, Complements of Cosmology in the Theory of Ether,PIRT 2004.